\documentclass[twocolumn,aps,pre,showpacs,amssymb,floatfix]{revtex4}
\usepackage{graphicx}
\usepackage{amsmath}
\usepackage{amssymb}

	\newcommand{\beq}{\begin{equation}}
	\newcommand{\eeq}{\end{equation}}
	\newcommand{\beqa}{\begin{eqnarray}}
	\newcommand{\eeqa}{\end{eqnarray}}
\def\barr{\begin{array}}
\def\earr{\end{array}}
\def\gamdot{\dot{\gamma}} 
\def\qzero{q_0}

\begin{document}

\title{Shear-driven size segregation of granular materials: modeling and experiment}

\author{Lindsay B.~H. May,$^1$ Laura A. Golick,$^2$ Katherine C. Phillips,$^2$
Michael Shearer,$^1$ Karen E. Daniels$^2$}

\affiliation{$^1$Department of Mathematics and $^2$Department of Physics, North Carolina State University, Raleigh, NC, USA}
\date{November 15, 2009}

\begin{abstract}
Granular materials segregate by size under shear, and the ability to quantitatively predict the time required to achieve complete segregation is a key test of our understanding of the segregation process. In this paper, we apply the Gray-Thornton model of segregation (developed for linear shear profiles) to a granular flow with an exponential profile, and evaluate its ability to describe the observed segregation dynamics. Our experiment is conducted in an annular Couette cell with a moving lower boundary. The granular material is initially prepared in an unstable configuration with a layer of small particles above a layer of large particles. Under shear, the sample mixes and then re-segregates so that the large particles are located in the top half of the system in the final state. During this segregation process, we measure the velocity profile and use the resulting exponential fit as input parameters to the model. To make a direct comparison between the continuum model and the observed segregation dynamics, we locally map the measured height of the experimental sample (which indicates the degree of segregation) to the local packing density. We observe that the model successfully captures the presence of a fast mixing process and relatively slower re-segregation process, but the model predicts a finite re-segregation time, while in the experiment re-segregation occurs only exponentially in time.
\end{abstract}


\maketitle

\section{Introduction} 

Granular materials have long been known to segregate by size, shape, density, and other material properties, whether driven by shear or vibration \citep{Ottino-2000-MSG, Kudrolli-2004-SSV, Schroter-2006-MSS}. The case of size-segregation in shear flow is particularly important, as it arises in such diverse situations as industrial chute flows, rock avalanches, and rotating tumblers \citep{Drahun-1983-MFS, Iverson-2001-NVG, Taberlet-2006-AST, Zik-1994-RIS, Khakhar-1997-RSG}. Under such shear, large particles typically rise and small particles descend. The dominant mechanism in such cases is thought to be percolation-based, where the granular flow acts as a sieve through which the small particles preferentially fall, but arguments based on kinetic theory have been proposed as well \citep{Hsiau-1996-GTD, Yoon-2006-IDS}. A number of groups \citep{Bridgwater-1985-PMS, Savage-1988-PSS, Gray-2005-TPS} have developed continuum models to describe vertical size-segregation within the constant shear-rate flows typical of free-surface avalanches. The aim of this paper is to apply a recent modification \citep{May-2009-SCL} of the Gray-Thornton model \citep{Gray-2005-TPS} to a boundary-driven granular system where the shear rate is a nonlinear function of depth. We quantitatively investigate the model's ability to capture experimentally-observed segregation dynamics and the associated changes in overall packing density.

\begin{figure}[b]
\centerline{\includegraphics[width=\columnwidth]{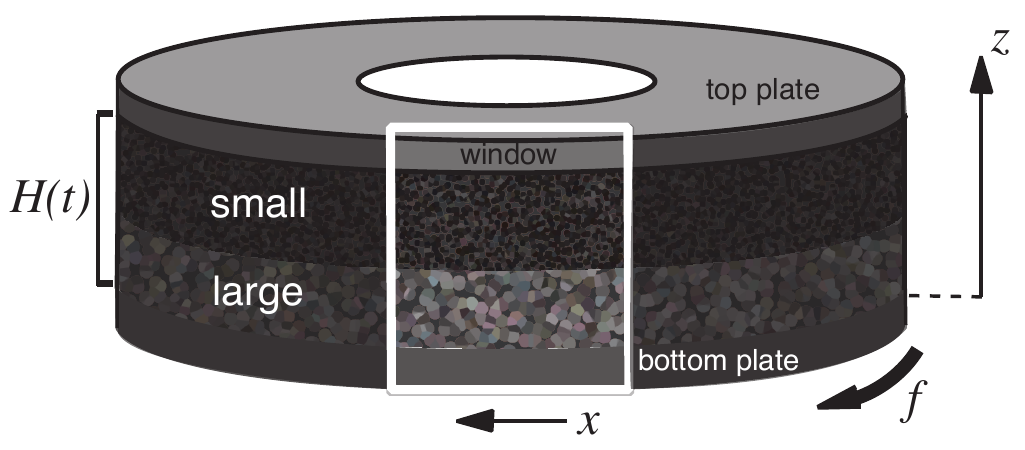}}
\caption{Schematic of experimental apparatus (not to scale) showing initial configuration and coordinate system. 
\label{schematic}}
\end{figure}

We perform experiments on binary granular mixtures of spherical particles confined in an annular Couette geometry (see Fig.~\ref{schematic}). The system starts in an unstably-stratified state with small particles over large particles and progresses to the reverse, stable, configuration as the  lower boundary is rotated. Key advantages of this geometry are the ability to run continuously without the need to feed material, as would be the case for chute flows, and the ability to start and end the experiment in well-defined states. As is commonly the case for sheared granular materials \citep{Midi-2004-DGF}, we observe that the velocity profile is not a linear function of depth as assumed in \citep{Savage-1988-PSS, Gray-2005-TPS, Gray-2006-TDS}, but instead decays exponentially away from the shearing surface. Correspondingly, the segregation rate is not uniform, but is higher near the bottom rotating plate. We use high-speed digital imaging and particle-tracking to measure the velocity profile $u(z)$ in the experiment and use the resulting fit to an exponential form as input to the model.

Kinetic sieving models of segregation \citep{Bridgwater-1985-PMS, Savage-1988-PSS, Gray-2005-TPS} are based on the notion that vertical size-segregation is principally caused by small particles preferentially falling into gaps created by the relative motion (shear) of the particles beneath them. Secondarily, the large particles are forced upward in a process sometimes called squeeze expulsion. In particular, \citet{Bridgwater-1985-PMS} argue that the vertical velocity of small particles due to segregation (the {\itshape segregation velocity}) should be proportional to the shear rate $\gamdot \equiv | \partial u / \partial z |$ as well as some function of the concentration $\phi$ of small particles. 
In the context of chute flow and avalanches, both \citet{Savage-1988-PSS} and \citet{Gray-2005-TPS} assume a segregation velocity which is additionally proportional to the concentration of large particles, $(1-\phi)$, since large particles provide the gaps into which the small particles fall. The choice of a constant of proportionality (known as the {\itshape segregation rate}) comes from an assumption of constant shear rate, whereas for many important granular systems it instead falls off exponentially \citep{Midi-2004-DGF}.

Since the local shear rate is set by the local horizontal velocity, we measure $u(z)$ directly from the experiment.  
In our measurements of the horizontal velocity profile, described in \S\ref{experiment}, we observe that it falls off exponentially away from the shearing surface, is the same for both particle sizes, and is approximately time-independent except for an initial transient. Using the steady-state velocity profile $u(z)$ to determine the shear rate, we calculate an exact solution, described in \citet{May-2009-SCL} and summarized in \S\ref{model}.
The concentration of small particles, calculated as a function of depth and time, predicts a rarefaction wave that mixes the particles over short time. As small particles reach the bottom plate, a growing layer of small particles is formed, bounded by a shock wave (a discontinuity in concentration) propagating upwards. Correspondingly, large particles eventually reach the top of the annulus, and a second shock propagates downward. The two shocks meet in finite time, after which the material is completely re-segregated. 

While it is not possible to monitor the local concentration field $\phi(z,t)$ within the experiment, we measure the progression of the mixing and segregation processes via the compaction and subsequent re-expansion of the aggregate. To relate the dynamics of the exact solution to the experiment, we postulate that the local concentration $\phi$ of small particles determines the local packing density $\rho$, as described for static packings in \citet{Kristiansen-2005-SRP}. This {\it concentration map} $\rho(\phi)$ allows us to model the change in the measured height $H(t)$ of the aggregate as a function of time. We compare this {\itshape proxy height} $\tilde H(t)$ to the time evolution of the experimentally measured height in \S\ref{compare}.

In \S\ref{disc}, we evaluate the successes and failures of the model. To make a quantitative comparison, we set two of the three free model parameters from the observed system height and the magnitude of the transient compaction. The third parameter is the constant of proportionality between shear rate and segregation rate, and sets the overall duration of the process. This segregation rate has previously been observed to vary with such parameters as particle size-ratio and confining pressure \citep{Golick-2009-MSR}.
We find that the model successfully captures the existence of a fast mixing process followed by a slower re-segregation process. However, the model re-segregates in finite time and does not replicate the observed behavior from the experiment, where a completely re-segregated state is approached only exponentially in time. 

\section{Experiment \label{experiment}} 

The experimental apparatus is an annular Couette cell filled with a bidisperse mixture of spherical glass particles confined by cylindrical walls at inner radius $(25.5 \pm 0.1)$ cm and outer radius $(29.3 \pm 0.1)$ cm. We apply shear via a circular bottom plate rotating at a constant frequency $f = 49 \pm 0.5$ mHz, approximately 3 rpm. 
A heavy top plate sits within the annular gap and is free to move vertically, but is partially suspended by springs to reduce the pressure it applies. The compressive force $P$ applied to the particles is $(0.36 \pm 0.008) \, mg$, where $mg$ is the total weight of the particles and the variation in force is due to the stretching of the springs. The top and bottom plates have a rubberized surface to increase friction with the particles, while the stationary cylindrical side walls are constructed of aluminum. 

\begin{figure}
\centerline{\includegraphics[width=\columnwidth]{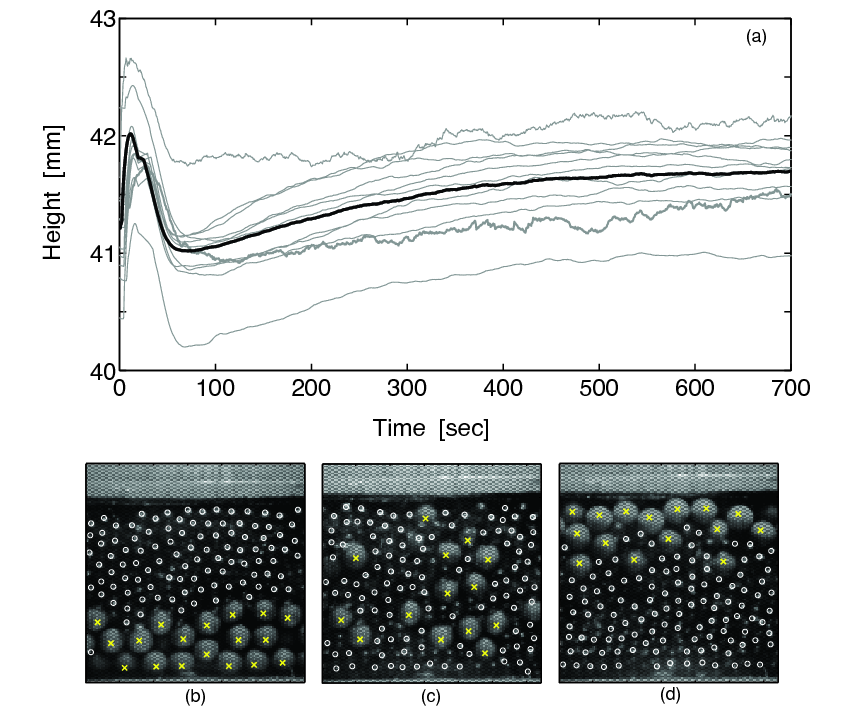}}
\caption{(a) $H(t)$ for 11 experimental runs (gray lines), with thicker line representing the run for which we took movies and measured the velocity profile. The thick black line is the average of the 11 runs, used for comparison with the model in Fig.~\ref{heightcompare}. Positions of particles ($\circ$ indicates small, $\times$ indicates large) overlaid on images taken through the window in the outer wall (b) initially, $t = 0$~sec, (c) in a mixed state, $ t = 127$~sec, and (d) in the re-segregated state, $t = 1820$~sec. \label{timeandsnapshots} }
\end{figure}

At the start of each experiment, we prepare a flat layer of 2.0 kg of large particles (diameter $d_L= 6$~mm) at the bottom of the annulus, followed by a flat layer of 2.0 kg of small particles ($d_S = 3$~mm). This configuration is shown schematically in Fig.~\ref{schematic}. We consolidate the layers by compression prior to the beginning of the run, with the average initial height $H(0) = 41.2$~mm. 

Fig.~\ref{timeandsnapshots} shows the height $H(t)$ of the top cell as a function of time for several different runs under the nominally identical conditions. While the variability from run to run is considerable (and typical of granular materials), the important features are common to all of the runs. From the initial normally-graded configuration, the applied shear causes the lowest (large) particles to move horizontally due to friction with the rubberized bottom plate. As shear begins, the material must dilate in order to deform (Reynolds dilatancy). After this initial rapid expansion, $H(t)$ provides information about the progression of mixing/segregation in the aggregate. Because the mixing process causes small particles to partially fill the voids between the large particles, the total height of the aggregate decreases. For all runs, the height reaches a minimum $H_{min}$ at $t\approx 80$~sec, with $\Delta H \approx 1$~mm. Further shearing serves to re-segregate the particles, with the large particles ending at the top of the cell and the small particles at the bottom. During this process, $H$ rises back to a height greater than its initial value. This re-segregation process would in theory continue until there were no longer a mixture of particles in the central region. In practice, however, the particles do not fully re-segregate and a few large particles remain within the lower layer, as can be seen in Fig.~\ref{timeandsnapshots}d. 


To measure the velocity of the particles, we observe the outer layer through a window of approximate width $10$~cm, using a digital video camera operating at 450 Hz. We recorded images during three time intervals, each with an approximate duration of $10$ minutes (a total of around $10^6$ images), separated by intervals of similar duration during which images were transferred from the camera to the computer. The system reaches a re-segregated state after approximately $t_f=700$~seconds. 

From each image, we first identify the center of each particle by convolution with a circular kernel chosen to match either $d_S$ or $d_L$. We perform the convolution twice (once for each particle-diameter) and then screen for mis-detections and double-detections. Fig.~\ref{timeandsnapshots} shows particle centers for (b) particles in the initial, normally-graded configuration, (c) the mixed state, and (d) the re-segregated, inversely-graded state. Because the particular configuration visualized at the wall only measures the state of a small portion of the system, we use the images only for calculating the velocity profile, and use $H(t)$ to probe the average degree of mixing/segregation. 

\begin{figure}
\centering
\includegraphics[width=\columnwidth]{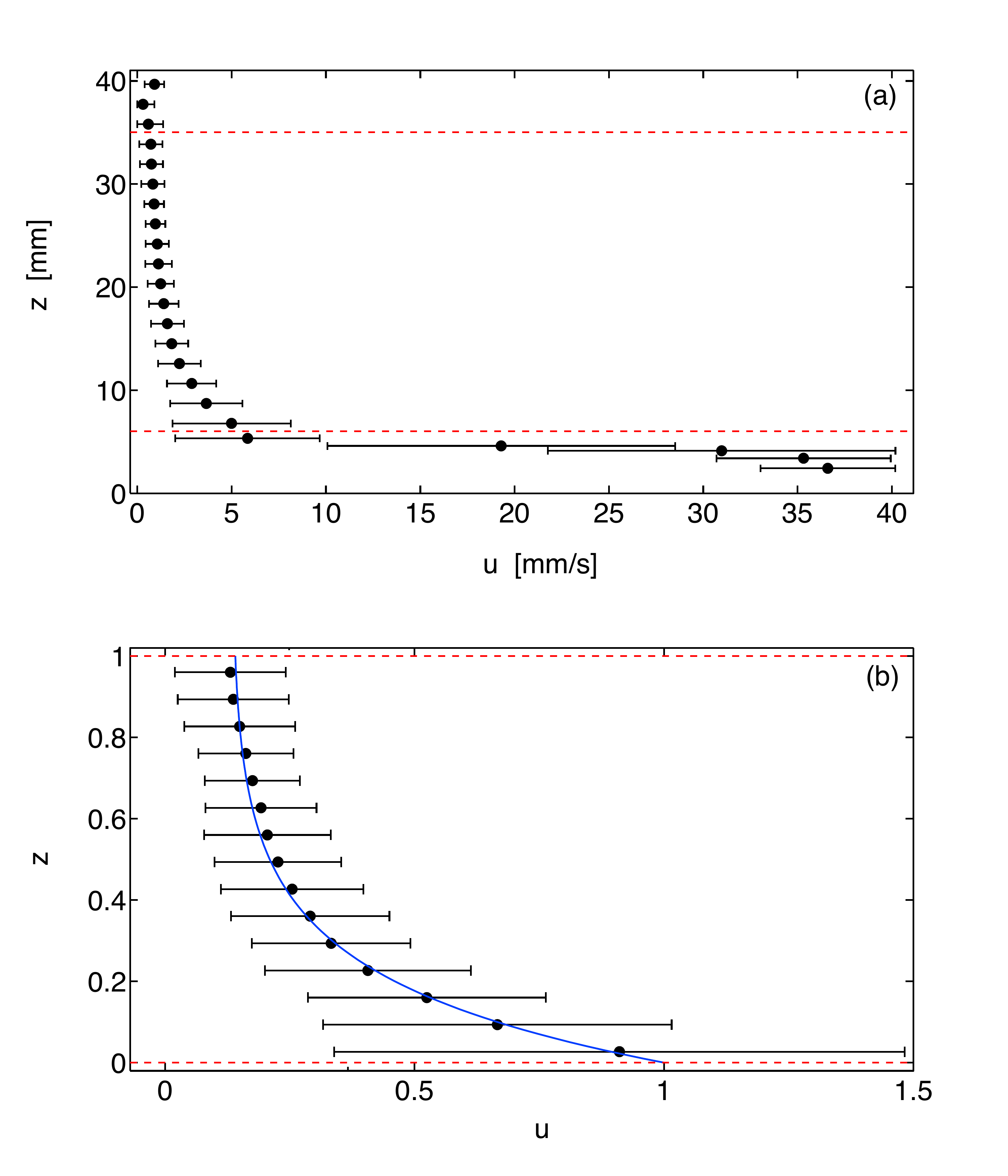}
\caption{Measured velocity profile $u(z_i)$ ($\bullet$) for (a) full cell height, with boundary layers of thickness $d_L$  above and below the dashed horizontal lines and
(b) within the central region, scaled so that $z \in [0,1]$. The velocities are scaled by $u = 5.5$~mm/sec so that $u(0) = 1$. The solid curve is the fit to Eq.~(\ref{velocity1}). Bars represent width of the velocity distribution at half the height of the peak.
\label{VP}}
\end{figure}

To characterize the shear, we are primarily interested in the average horizontal velocity $u$ and how it depends on depth, time, and particle size. We assemble particle trajectories from the list of particle positions associated with each image, considering each of the two sizes separately. For each trajectory, we calculate the instantaneous velocity of the particle from the slope of a linear fit over a duration appropriate to the average speed of the layer. This analysis was repeated separately for each of the three 10 minute intervals to check for time-independence; we include all images with $t < 3100$ seconds in order to improve our statistics. We observed a steady-state velocity profile for all three intervals (shown in Fig.~\ref{VP}), after an initial $\approx 37$~sec transient which is excluded from the analysis. We observered that $u(z)$ was approximately the same for both large and small particles.

\begin{figure}
\centering
\includegraphics[width=\columnwidth]{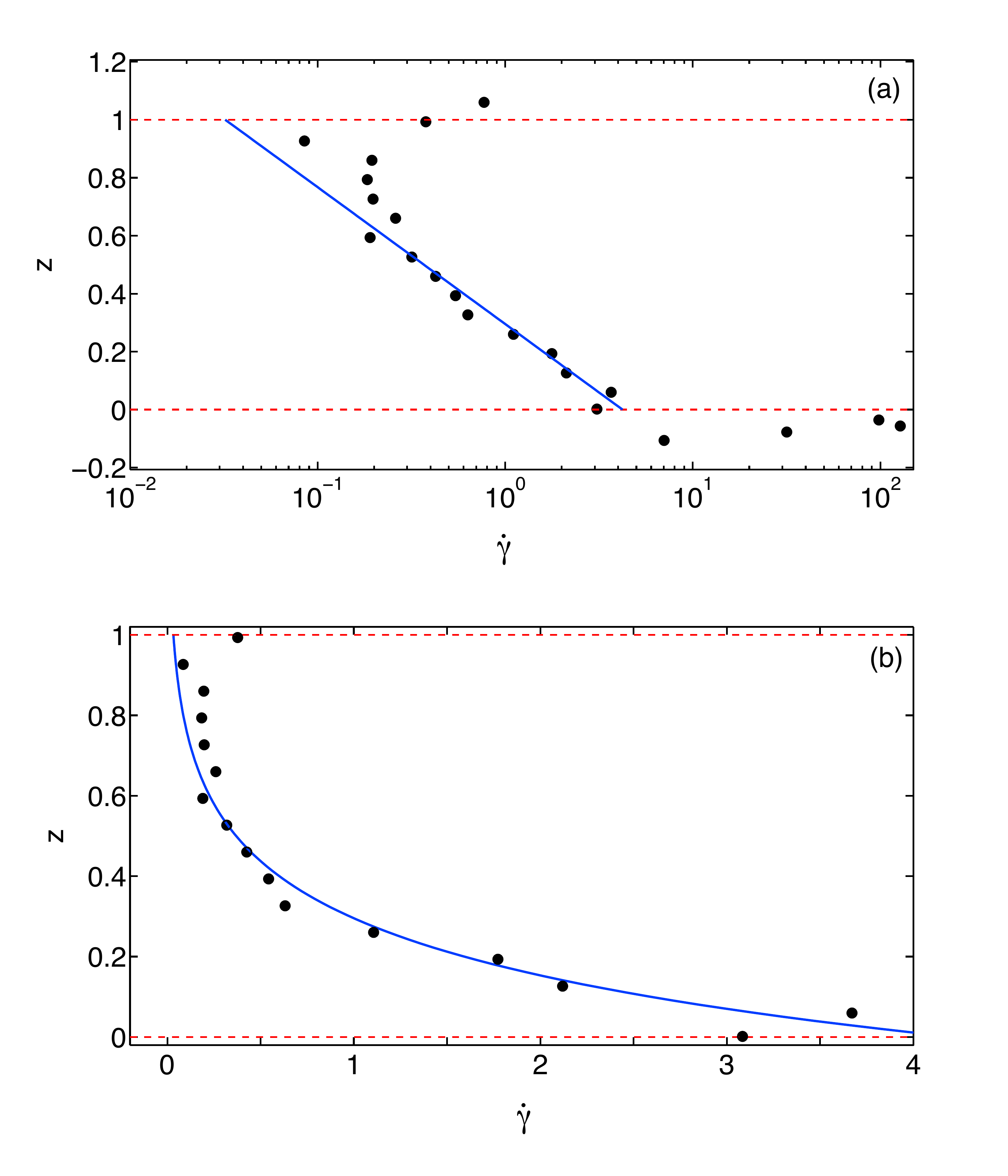}
\caption{Dimensionless shear rate $\dot \gamma = |du/dz|$ ($\bullet$) for (a) full cell height, with boundary layers above and below the dashed horizontal lines and (b) within the region $z \in [0,1]$. Solid line is the fit from Fig.~\ref{VP}, plotted as Eq.~(\ref{vprimeofz}).}
\label{SR}
\end{figure}

We measure the average horizontal velocity as a function of depth by dividing the ensemble of trajectories into a discrete set of bins centered at positions ${z_i}$. Within each bin, we plot a probability distribution of the velocities and fit a parabola to the peak. This maximally-likely velocity, $u(z_i)$, is plotted in Fig.~\ref{VP}, with the width of the distribution at half the height of the peak represented by the horizontal bars for each $z_i$. Note that the velocity fluctuations (``granular temperature'') are largest within the shear band, and that thermal gradients have been associated with segregation in kinetic theory models \citep{Hsiau-1996-GTD, Yoon-2006-IDS}.

Since the segregation is driven by shear in our model, we calculate the experimental shear rate $\gamdot(z) \equiv |\partial u / \partial z|$ from velocity data by finite-differences, shown in Fig.~\ref{SR}. Note that the bottom and top layers, of depth approximately $d_L$, show dramatically higher shear rates than the bulk. In the figures, we mark the boundaries between the bulk and the two layers by horizontal dashed lines. In order to model the long-term mixing and re-segregation, we focus on the central region of the flow and scale the $z$ axis so that $z=0$ at the bottom of this region and $z=1$ at the top. Within this region, we observe that the velocity profile is well-described by an exponential of the form 
\beq
u(z) = u_0 e^{-z/\lambda} + c,
\label{velocity1}
\eeq
where $\lambda$ is related to the width of the shear band, $u_0 + c$ is the velocity at $z=0$, and $c$ is a constant representing solid-body rotation. The fit to this form is plotted as a solid line in Fig.~\ref{VP}b, scaled so that $u(0) = 1$. The fit parameters are $\lambda = 0.205$, $u_0 = 0.866$, and $c = 0.134$.

To obtain $\gamdot(z)$ for use in the model, we differentiate $u(z)$ in Eq.~(\ref{velocity1}) of $u(z)$ and find 
 \beq
 \gamdot(z) =\frac{u_0 }{ \lambda} \, e^{-z/\lambda}
 \label{vprimeofz}
 \eeq
This function is plotted in Fig.~\ref{SR}b for comparison with the experimentally-determined profile obtained by finite-differencing. In \S\ref{model}, we will relate the local segregation rate to the parameters $u_0$ and $\lambda$.

\section{Segregation Model} \label{model} 

To simulate the segregation and mixing in the experiment, we begin with an equation for the conservation of mass and impose our modeling assumptions to mimic conditions in the experiment. In the model, we consider the kinetic sieving mechanism but not diffusion. The conservation of mass equation is \cite{Gray-2005-TPS}:
 \beq\label{mascons}
 \frac{\partial\phi}{\partial t} + \frac{\partial}{\partial x}(\phi u) 
 + \frac{\partial}{\partial y}(\phi v) +\frac{\partial}{\partial z}(\phi w)
 + \frac{\partial}{\partial z}\big(q(z) \,  \phi (\phi-1)\big) = 0,
 \eeq
where $\phi=\phi(z,t)$ is the concentration of small particles at position $z$ and time $t$. The velocity $(u,v,w)$ is the bulk velocity of small particles. Kinetic sieving is modeled by a modification $w_s=q(\phi-1)$ to the vertical component $w$ of bulk velocity, in which $q = q(z)$ is the local segregation rate.

We assume that the components $v,w$ of the bulk velocity (in the $y, z$ directions, respectively) are negligible. That is, we assume there is essentially no motion across the flow, and that the vertical component of velocity of small particles is dominated by the effect of segregation. In the annular Couette geometry of the experiment, the flow is uniform in the $x$-direction (the angular direction); we assume $u = u(z)$ is independent of horizontal position $(x,y)$ and time $t$. Finally, we assume that the segregation rate $q(z)$ depends only on the vertical variable $z$ and is proportional to the shear rate $\gamdot(z)$:
\beq\label{segrate1}
q(z)=s\gamdot(z)=\qzero  e^{-z/\lambda}, \quad \qzero \equiv s\frac{u_0 }{ \lambda}.
\eeq
The constant of proportionality $s$ sets the time scale in the model and represents the rate at which mixing and segregation processes occur. In previous experiments, we observed that the timescale of both the mixing and segregation processes is a function of particle size ratio and confining pressure \cite{Golick-2009-MSR}. 

With these assumptions neglecting small velocities, the model (\ref{mascons}) reduces to the scalar conservation law 
 \beq 
\frac{\partial \phi}{\partial t} + \frac{\partial }{\partial z} \big( q(z) \, \phi(\phi-1)\big)=0.
 \label{pde1}
 \eeq 
We set an initial condition corresponding to the beginning experimental configuration of a layer of large particles above a layer of small particles:
 \beq
 \phi(z,0)=\left\{\barr{ll}
 0, & 0< z<z_0, \\[10pt]
 1, \quad&z_0<z<1,
 \earr\right.
 \label{init}
 \eeq
with $z_0 = \frac{1}{2}$. Boundary conditions 
 \beq
 \phi(0,t)=1, \quad \phi(1,t)=0
 \label{bc}
 \eeq
ensure that there is no flux of the particles through the upper and lower boundaries.

In \citet{May-2009-SCL}, we constructed the solution of Eq.~(\ref{pde1}-\ref{bc}) using characteristics and shock waves. The early time solution is a rarefaction wave centered at $z=z_0,\ t=0$, that represents the initial mixing, in which $\phi$ varies continuously from $\phi=0$ to $\phi=1$. The solution in the rarefaction fan  is characterized by a pair of equations
\begin{subequations}
\beq\label{phi_rare}
\phi(z,t)=-\sigma_0 t \phi_o(1-\phi_o)e^{-z_0/\lambda}+\phi_o,
\eeq
\beq\label{quad1}
 e^{z/\lambda}-e^{z_0/\lambda}= - 
\phi_o(1-\phi_o)e^{-z_0/\lambda}(\sigma_0t)^2+(2\phi_o-1)\sigma_0t,
\eeq
\end{subequations}
where $\sigma_0=\qzero/\lambda$.  In these equations, $0\leq\phi_0\leq 1$ labels a specific characteristic at $t=0, z=z_0;$ $\phi_0(z,t)$ can be found by solving the quadratic equation (\ref{quad1}), and choosing the relevant solution. Then Eq.~(\ref{phi_rare}) is an explicit formula for $\phi(z,t).$

\begin{figure}
\centerline{\includegraphics[width=\columnwidth]{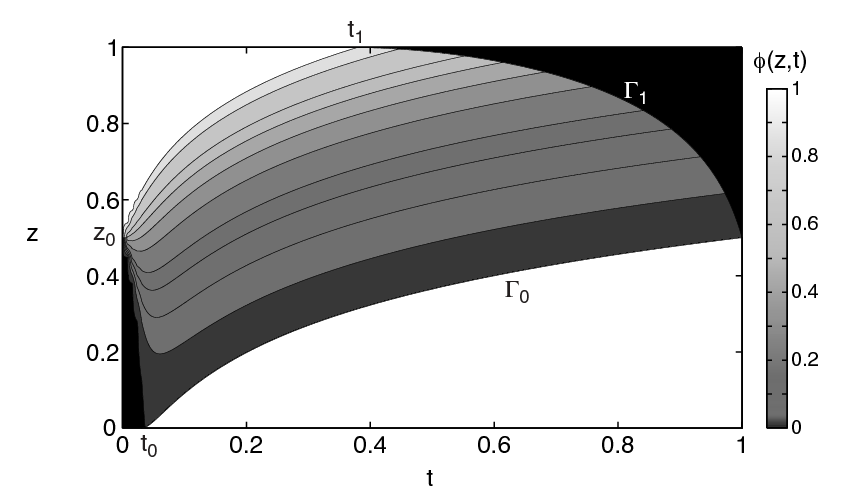}}
\caption{Numerical solution of the initial value problem Eq.~(\ref{pde1} - \ref{bc}), with parameters $u_0 = 0.866, \lambda= 0.205 , z_0=\frac{1}{2}$. For this plot, we set the segregation rate $s = 16.2$~s$^{-1}$ so that $t_f = 1$ (non-dimensional) at the final, fully-segregated state. \label{solution}}
\end{figure}

The rarefaction reaches the bottom and top boundaries when the characteristics $\phi=1, \phi=0$ reach $z=0, z=1$, respectively. This corresponds to the first small particle reaching the bottom plate and the first large particle reaching the top plate. These events occur at times
 \beq
 t_0 = \frac{\lambda}{\qzero }(e^{z_0/\lambda}-1), \qquad 
 t_1 = \frac{\lambda}{\qzero }(e^{1/\lambda}-e^{z_0/\lambda}),
 \eeq
respectively. Subsequently a layer of small particles grows from $z=0$, and a layer of large particles grows from $z=1$. The interfaces $z=\Gamma_0(t), z=\Gamma_1(t)$ between the rarefaction and these layers are shock wave solutions of the conservation law (\ref{pde1}), and  consequently evolve according to the Rankine-Hugoniot condition, which for Eq.~(\ref{pde1}) is a differential equation for each shock:
\begin{subequations}
\beq
\Gamma_0'(t) =q(\Gamma_0(t))\phi(\Gamma_0(t),t), \quad t>t_0, \ \Gamma_0(t_0)=0,
\eeq
\beq
\Gamma_1'(t) = q(\Gamma_1(t))(\phi(\Gamma_1(t),t)-1), \quad t>t_1, \ \Gamma_1(t_1)=1.
\eeq
\label{gammas}
\end{subequations}
When these shocks meet, re-segregation is complete, and the solution consists of a stationary shock separating the upper layer of large particles from the lower layer of small particles. By mass conservation, the position of the interface is $z=1-z_0$. A contour plot of the solution corresponding to parameter values calculated from the experimental data is shown in Fig.~\ref{solution}. 

\section{Comparison of Model and Experiment}\label{compare}

In the experiment, the height $H(t)$ initially increases due to Reynolds dilatancy, then decreases sharply as the material mixes (see Fig.~\ref{timeandsnapshots}a). Subsequently, $H(t)$ increases at a slower rate as the material re-segregates. In this section, we relate the mixing and re-segregation stages in the height evolution to predictions of the model (\ref{pde1}), shown in Fig.~\ref{solution}. 

The solution $\phi(z,t)$ of the model is specified at each point $z$ in the fixed domain $0\leq z\leq 1$, and at each time $t$. However, the physical domain varies with time, due to the changes in height $H(t)$. To relate the evolution of $\phi(z,t)$ to $H(t)$, we calculate a proxy height that takes into account the different packing densities of the material at different concentrations $\phi$ of small particles. The {\it local packing density} $\rho$ is the ratio of the solid local volume occupied by the mixture of particles to the total local volume, i.e., including void space. The total volume of the annular region is the product of the cross sectional area $A$  and the position $H$ of the top plate. For a local solid volume of particles $dV_p$ across a cross section of the annulus, in a horizontal layer of small (physical) thickness $dh$, we write 
\begin{equation}
\rho = \frac{dV_{p}}{A \, dh}.
\end{equation}
We will assume that the local packing density $\rho$ depends only on the local concentration $\phi(z,t)$. That is, we take $\rho=\rho(\phi)$, independent of other factors such as the shear rate. Thus, for each $dz$ (the height element in the model domain, $0\leq z \leq 1$), there will be a local contribution to $H(t)$ which depends on the local $\phi$. In order to convert from the model domain to the physical domain, we integrate to obtain a proxy height $\tilde{H}(t)$:
\begin{equation}
\tilde{H}(t) = \int^{\tilde{H}(t)}_0{dh} = \frac{V_p}{A}\int_0^1{\frac{dz}{\rho\left(\phi(z,t)\right)}}.
\label{proxyformula}
\end{equation}
where $V_p$ is the total solid volume of the granular material. We note that in Eq.~(\ref{proxyformula}), $\frac{V_p}{A}$ has units of length, and the integral over $dz$ is nondimensional. Another way to interpret this equation is that the integral calculates the height of the sample in the mathematical model and the coefficient $\frac{V_p}{A}$ relates the  model height $\tilde{H}(t)$ to the physical height $H(t)$.


To determine an appropriate function $\rho(\phi)$, we first consider the  monodisperse cases in which $\phi = 0$ or $\phi =1$. In these cases, the lower limit for $\rho$ is known as random loose packing, ($\rho^{RLP} \approx 0.55$) and the upper limit as random close packing ($\rho^{RCP} \approx 0.64$) \citep{Bernal-1964-SL,Onoda-1990-RLP}. However, for bidisperse mixtures, the packing density depends on the relative composition of the mixture. Since small particles can fit within the spaces between large particles, the packing density is larger for a bidisperse mixture than for a monodisperse sample. Indeed, the maximum packing density for a bidisperse mixture with this size ratio has been observed to be around $\rho^{RCP, bi} = 0.67$ to $0.69$, depending on the method used \citep{Clarke-1987-NSD, Kristiansen-2005-SRP, Biazzo-2009-TAP}. 

Data from several numerical and experimental studies \citep{Rassouly-1999-PDP, Kristiansen-2005-SRP, Lochmann-2006-SAR} of the random close packing of a static bidisperse mixture of spheres supports an approximately triangular shape for the concentration map $\rho(\phi)$, with a maximum at $\phi_c = 0.275$. The function
\begin{eqnarray}\label{KristiansenFormula}
 \rho\left(\phi(z,t)\right)= \left\{\begin{array}{lll}
\rho_{min} + \Delta\rho\frac{\phi}{\phi_c}, &\quad & \phi\leq \phi_c,\\[10pt]
\rho_{min} + \Delta\rho\frac{1-\phi}{1-\phi_c}, &\quad & \phi > \phi_c,
 \end{array}
 \right.
 \end{eqnarray}
falls off, by an amount $\Delta \rho$, to minimum packing density $\rho_{min}$ at a $\phi=0,1$. \citet{Kristiansen-2005-SRP} report a value of $\rho_{min} = 0.628$, which is below both the monodisperse and bidisperse $\rho^{RCP}$.

Note, however, that there is a significant difference between the conditions under which these studies determined the packing density, and the conditions of our experiments: our measurements take place in a sheared system rather than a static one. The packing density for sheared granular materials is typically less than for static packings, due to Reynolds dilatency. However, we are not aware of any measurements, analogous to those reported in \citet{Kristiansen-2005-SRP}, for $\rho(\phi)$ in sheared granular materials. (One can imagine that segregation in fact makes such measurements rather challenging.) Therefore, we must estimate the values of $\rho_{min}$ and $\Delta\rho$ in the concentration map (\ref{KristiansenFormula}) for our bidisperse mixture of granular materials under shear. We expect that the value of $\rho_{min}$ may be less than the value found in \citet{Kristiansen-2005-SRP} due to the shear, but is probably not less than the random loose packing limit for monodisperse spheres, $\rho^{RLP}$. Therefore, we consider $0.55 < \rho_{min} < 0.64$ as a reasonable range of values. Typical reported values of $\Delta\rho$ range from $0.025$ to $0.063$ \citep{Kristiansen-2005-SRP}. Since $\rho^{RCP,bi}$ provides a larger upper limit for bidisperse mixtures, we consider a large range of values, $0.02 < \Delta\rho < 0.14$.

\begin{figure}
\centerline{\includegraphics[width=\columnwidth]{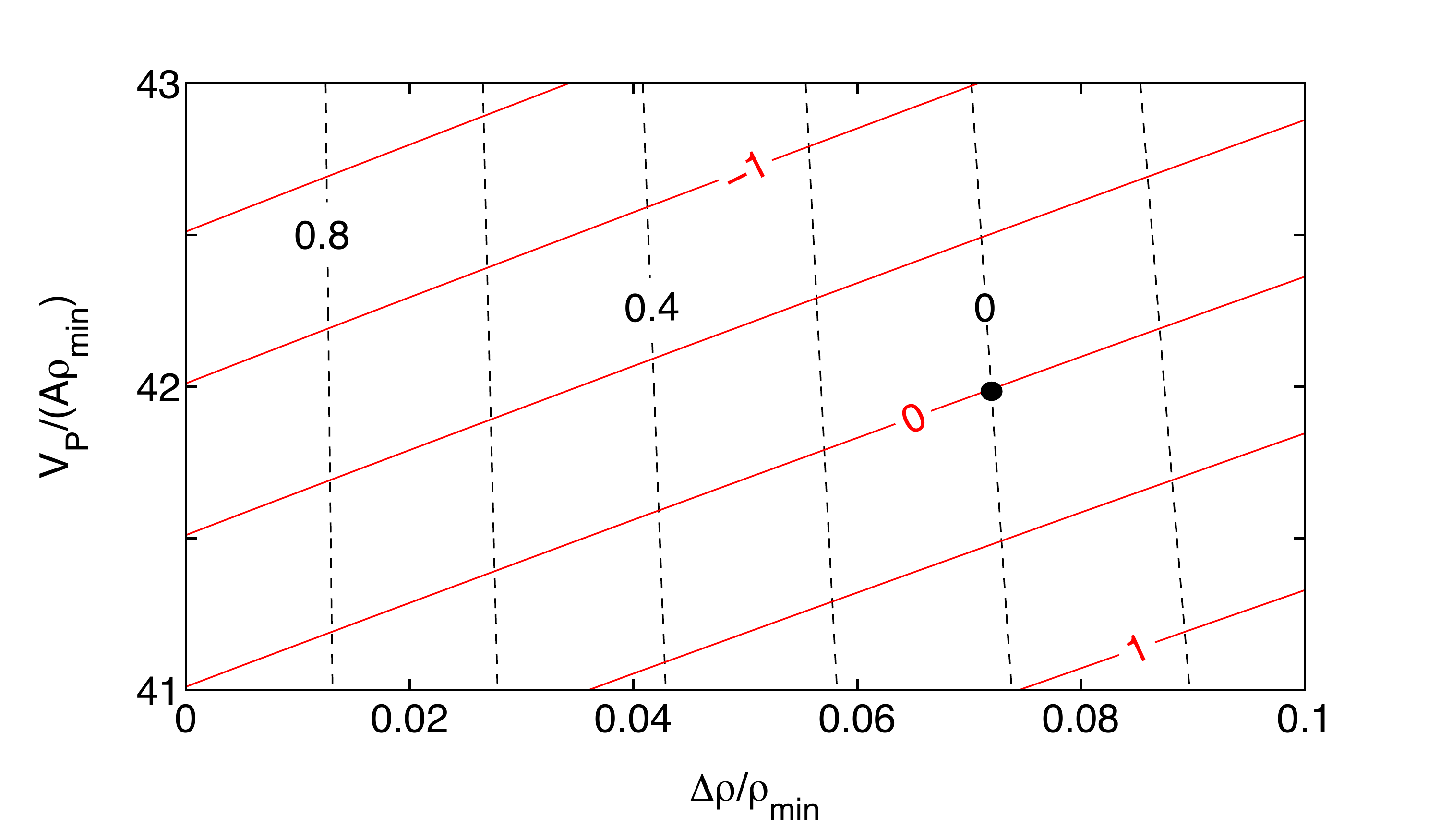}}
\caption{Contour plot of $H_{min} - \tilde{H}_{min}$ (solid lines) and $\Delta H - \Delta \tilde{H}$ (dashed). Symbol $\bullet$ marks the intersection of the two zero contours at $V_p/(A\rho_{min}) = 42.0$~mm and $\Delta\rho/\rho_{min} = 0.072$.
\label{paramfind}}
\end{figure}

Substituting Eq.~(\ref{KristiansenFormula}) into (\ref{proxyformula}), we observe that there are only two free parameters: $\frac{V_p}{A\rho_{min}}$ sets the overall height of the system and $\frac{\Delta\rho}{\rho_{min}}$ sets the amount of compaction/expansion.  From measurements of apparatus dimensions and particle sizes, we find $\frac{V_p}{A} = 24.2 \pm 1.2$~mm. This range, together with the ranges for $\rho$ and $\Delta \rho$ given above, predict that we should consider parameters $37 \lesssim \frac{V_p}{A\rho_{min}} \lesssim 44$ and $0.07 \lesssim \frac{\Delta\rho}{\rho_{min}} \lesssim 0.26$. To select the values which best capture the compaction and expansion process, we perform the integral (\ref{proxyformula}) for pairs of $(\frac{V_p}{A\rho_{min}} , \frac{\Delta\rho}{\rho_{min}})$ values, and determine the resulting $\tilde{H}_{min}$ and $\Delta \tilde{H}$. In Fig.~\ref{paramfind}, we show contour plots of the difference between the experimentally-measured values and the proxy-calculated value over this full parameters range. The best parameter choice lies at the intersection of the two zero-contour lines. We find these values to be $\frac{V_p}{A\rho_{min}} = 42.0$ and $\frac{\Delta\rho}{\rho_{min}} = 0.072$. The first parameter is to be expected given the height of the system in Fig.~\ref{timeandsnapshots}, and the second is at the lower end of the expected range of values. 

\begin{figure}
\centerline{\includegraphics[width=\columnwidth]{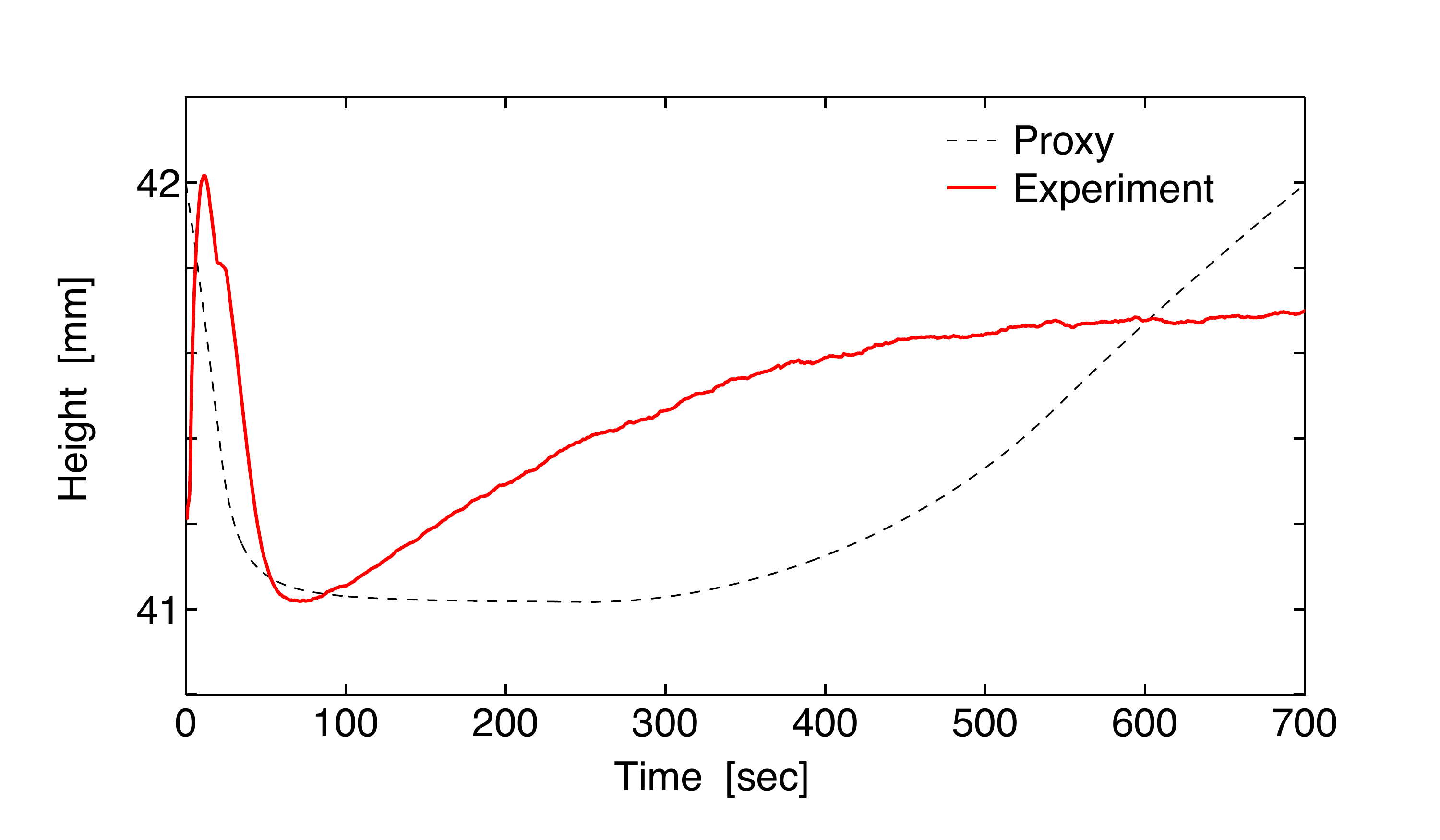}}
\caption{Comparison of the experimentally-measured $H(t)$ (solid line, red, see Fig.~\ref{timeandsnapshots}) and the calculated proxy height $\tilde{H}$ (dashed line, black) for the parameters given in Fig.~\ref{paramfind}. \label{heightcompare}}
\end{figure}

A third free parameter is the segregation rate $s$ in Eq.~(\ref{pde1}) that sets the time scale for the dynamics. From the solution of the initial value problem, we find that the choice $s = 0.023$ predicts the ending time for the solution equal to the experimentally-determined time $t_f = 700$~sec. Fig.~\ref{heightcompare} provides a direct comparison of the experiment and the model over this time interval. We observe that the proxy height decreases more rapidly than it climbs, in agreement with the experimentally observed $H(t)$; this feature arises without including separate mixing and segregation rates in the model. The micromechanical origins of this property remain to be investigated in future work. In addition, the rate at which it decreases is close to the observed rate. At later times, the shapes of $H(t)$ and ${\tilde H}(t)$ no longer agree, with ${\tilde H}(t)$ exhibiting both a flatter minimum and a faster re-segregation rate. It is particularly notable that the model solution predicts re-segregation in finite time, in contrast to the exponential approach to re-segregation observed in the experiment \citep{Golick-2009-MSR}. 

\section{Conclusions} \label{disc}

We have performed quantitative experiments to compare predictions for a recent extension \citep{May-2009-SCL} of the Gray-Thornton model of granular segregation \citep{Gray-2005-TPS} to  a flow in which the shear rate is non-constant. This comparison additionally requires the construction of a model connecting the local concentration of small and large particles to changes in the local packing density (the concentration map). When these two elements are combined, we can compare the temporal evolution of the height of the segregating system to a proxy height calculated from the continuum model. Several features of the experimentally-measured height dynamics $H(t)$ are well-captured by the proxy height: (1) a fast mixing timescale is followed by a much slower re-segregation timescale, (2) we can model the compaction and re-expansion process using reasonable parameter values in the concentration map, and (3) the slope of the height curve during the mixing phase is in approximate agreement with that of the model.

Some prominent features are missing from the model: there is no means to account for Reynolds dilatency, and the model segregates in finite time rather than exponentially-approaching a final, re-segregated state. This latter point of disagreement shows the limitation of using a continuum mixture-theory model for a discrete process, especially where the number of discrete objects (the particles) is comparatively small. In the experiment, it is easy to see how this finite-size effect takes over after the continuum model predicts complete segregation. Once most of the large particles have reached the upper layer, it becomes increasingly difficult for the few large particles remaining in the lower region containing mostly small particles to segregate to the upper layer. When large particles do not have large particle neighbors, the small particles cannot sieve between the large particles and thus move the large particles upward. In fact, some of the large particles never make it to the upper layer, even after runs of several days duration, instead remaining trapped among the small particles. 

Despite the differences between the experimental and proxy heights, the proxy height calculated from this simple continuum model captures the qualitative features of the experimental height time series. We have extended the Gray-Thornton model, which was developed for the case of an avalanche (uniform shear), to model nonuniform shear, as occurs in a one dimensional configuration in an annular Couette cell. It is noteworthy that a continuum model applied to a small scale granular system successfully captures the main phenomena of mixing and re-segregation. 

\section{Acknowledgments}
The authors are grateful to Nico Gray for discussions about the model, and to David Fallest and Dhrumil Patel for initial hardware development and experiments. This research was supported by the National Science Foundation under grant DMS-0604047, and the National Aeronautics and Space Agency under grant NNC04GB086.


\end{document}